\documentclass[aps,pra,reprint,amsmath,amssymb,nofootinbib]{revtex4-2}

\usepackage{graphicx}
\usepackage{bm}

\begin{document}

\title{Time Dilation and Center-of-Mass Normalization in the Page--Wootters
Formalism}

\author{J. S. Ardenghi}
\email{jsardenghi@gmail.com}
\affiliation{*IFISUR, Departamento de F\'{\i}sica (UNS-CONICET),
Avenida Alem 1253, Bah\'{\i}a Blanca, Buenos Aires, Argentina}

\date{\today}

\begin{abstract}
In this work, we study the emergence of relativistic effects in a
composite quantum clock within the Page-Wootters relational formulation of
quantum mechanics. We consider a system with internal and center-of-mass
degrees of freedom and analyze the conditioned evolution of the center-of-mass relative to the internal system treated as an internal clock. We show that the internal sector exhibits an effective time dilation. Retaining the full mass-energy structure of the composite
particle, a back-reaction of the internal energy on the center-of-mass sector
appears through an operator-valued normalization factor. As a direct consequence of the back-reaction, the conditioned center-of-mass dynamics is governed by a Schr\"{o}dinger equation that is non-local in the clock time, with an effective temporal non-locality of the
order of the Compton time of the composite system. Applying the formalism to
a center-of-mass momentum superposition, we find that the interferometric
visibility acquires a correction that is quadratic in the internal
energy, and therefore depends on the absolute distribution of clock energies
rather than only on the energy gaps that control the standard time-dilation
dephasing. Although this correction is parametrically small, it is a generic
feature of relativistic composite clocks and can, in principle, be isolated
through a differential visibility measurement comparing clocks with equal
transition frequency but different mean internal energies. 
\end{abstract}

\maketitle

\section{Introduction}

In standard quantum mechanics, time enters the Schr\"{o}dinger equation as
an external classical parameter. For a closed quantum system, however, such an external temporal
parameter should not be fundamental. This tension is important in canonical
approaches to gravity, where physical states are subject to a Hamiltonian
constraint and the Wheeler--DeWitt equation replaces the usual Schr\"{o}%
dinger evolution. In that setting, the conventional notion of time evolution
is not clear, leading to the well-known problem of time and motivating the
search for a relational notion of temporal evolution \cite{dewitt,isham,kuchar}.
A particularly appealing proposal in this direction is the Page--Wootters mechanism,
in which time is not introduced as an external parameter but as a quantum
degree of freedom associated with a clock subsystem \cite{paw}. The total
state of the closed system is taken to be stationary, while the evolution of the remaining degrees of freedom is recovered conditionally, by
correlating them with the readings of the clock. In this way, the usual
dynamical description emerges from entanglement and conditioning within a
timeless global state. Although this mechanism has been the subject of
criticism, most notably regarding the interpretation of conditional
probabilities, the status of propagators, and the ambiguity associated with
different clock choices, later developments have clarified and extended the
framework, showing that it can consistently reproduce quantum statistics and
provide a robust relational description of dynamics \cite{unruh,giova,marletto,trinity,hslrel,rijavec}. The mechanism has also been illustrated experimentally in an entangled two-photon system, in which the polarization of one photon serves as a clock for the other \cite{moreva}.
In relativistic quantum-clock models, the flow of time for a composite quantum system is encoded in the evolution of its internal degrees of freedom, and relativistic effects
couple these internal dynamics to the motion of the center of mass \cite{zych,pikovski,smith,zychbrukner2018,khandelwal,grochowski,castroruizpnas}.
In particular, proper-time differences can manifest
themselves through changes in the internal evolution of a clock, while
motion and gravity induce correlations between internal and external degrees
of freedom. This perspective suggests that a relational treatment of time is
especially natural for composite quantum systems, where clock behavior and
kinematics are inseparable \cite{rovelli}.
In this work, we use the Page-Wootters formalism to analyze a composite
quantum clock with internal and center-of-mass degrees of freedom. We show
that, once the internal dynamics is described relationally with respect to
the clock subsystem, an effective time dilation naturally emerges in the
internal sector. Moreover, when the description is expressed in the
center-of-mass frame, the conditioned state acquires a corresponding
normalization fixed by the center-of-mass dynamics which has experimental consequences in the visibility of interferometric setups. 
Signatures of proper time in the interferometric visibility are useful because
the visibility is a quantity that can be measured directly in matter-wave and
clock interferometry. Writing the effect of the internal energy as a correction
to the visibility thus links these relational-time ideas to something
accessible in the laboratory, and gives a concrete way to test how the internal
energy of a composite system affects its center-of-mass coherence.

This paper is organized as follows. In Sec.~\ref{sec:paw} we review the
Page--Wootters construction for a composite quantum system. In
Sec.~\ref{sec:td} we introduce the relativistic composite-clock Hamiltonian,
derive the conditioned dynamics and the emergent time dilation, obtain the
center-of-mass normalization factor, work out explicit clock models, and
analyze the resulting interferometric visibility. Finally, in
Sec.~\ref{sec:conc} we summarize our results and discuss their implications.

\section{The PaW mechanism}
\label{sec:paw}

The PaW\ mechanism consists in the partition of a Hilbert space $\mathcal{H}$
in two non-interacting parts $\mathcal{H=H}_{S}\otimes \mathcal{H}_{C}$,
where we can call $\mathcal{H}_{C}$ the Hilbert space of the clock and $
\mathcal{H}_{S}$ the Hilbert space of the system. The flow of time can be obtained from the entanglement between the $\mathcal{H}_{S}$ and $\mathcal{H}_{C}$ quantum degrees of freedom; indeed, this system-clock entanglement can itself be taken as a measure of the
distinguishable evolution undergone by the system \cite{boette}. An
internal observer in $\mathcal{H}_{S}$ sees a quantum state that evolves
with $t$ as $\left\vert \phi (t)\right\rangle $. To obtain the Schr\"{o}dinger
equation for $\left\vert \phi \right\rangle \in \mathcal{H}_{S}$, we can
consider that both quantum systems do not interact and the total Hamiltonian
can be written as
\begin{equation}
H=H_{S}\otimes I_{C}+I_{S}\otimes H_{C}  \label{1a}
\end{equation}%
where $H_{C}$ is the clock's Hamiltonian and $H_{S}$ is the Hamiltonian of
the remaining part and where $I_{C}$ and $I_{S}$ are identity operators in
the respective Hilbert spaces. More general total Hamiltonians with
interaction between the clock and the system can be considered (see \cite{alex})
but for simplicity we neglect them here. We can consider that there
are special vectors $\left\vert \psi \rangle \right\rangle $ that obey $%
H\left\vert \psi \rangle \right\rangle =0$ and we can expand it as
\begin{equation}
\left\vert \psi \rangle \right\rangle =\int\limits_{{}}^{{}}dt\chi
(t)\left\vert t\right\rangle \left\langle t\right\vert \otimes
\int\limits_{{}}^{{}}d\epsilon _{S}\left\vert \epsilon _{S}\right\rangle
\left\langle \epsilon _{S}\right\vert \psi \rangle \rangle  \label{2a}
\end{equation}%
where we have used the identity resolutions $I_{S}=\int\limits_{{}}^{{}}d%
\epsilon _{S}\left\vert \epsilon _{S}\right\rangle \left\langle \epsilon
_{S}\right\vert $ in $\mathcal{H}_{S}$ and $I_{C}=\int\limits_{{}}^{{}}dt%
\chi (t)\left\vert t\right\rangle \left\langle t\right\vert $ the identity
resolution in $\mathcal{H}_{C}$, where $\left\vert \epsilon
_{S}\right\rangle $ and $\left\vert t\right\rangle $ are some basis that
expand $\mathcal{H}_{S}$ and $\mathcal{H}_{C}$ respectively. The quantum
states $\left\vert t\right\rangle $ are eigenstates of the time operator $T$
that obey $\left[ T,H_{C}\right] =i\hbar I_{C}$. The $\chi (t)$ is a
square-integrable function that guarantees the normalization condition for $%
\left\vert \psi \rangle \right\rangle $ (see Eq.~(23) of Ref.~\cite{giova}). By
using that%
\begin{equation}
e^{-\frac{i}{\hbar }I_{S}\otimes H_{C}t^{\prime }}I_{S}\otimes Te^{\frac{i}{%
\hbar }I_{S}\otimes H_{C}t^{\prime }}=I_{S}\otimes T-t^{\prime }I_{S}\otimes
I_{C}  \label{2a1}
\end{equation}%
and using that the clock operator can be written as $T=\int%
\limits_{{}}^{{}}dt\chi (t)t\left\vert t\right\rangle \left\langle
t\right\vert $ and replacing this in the last equation we obtain
\begin{widetext}
\begin{equation}
e^{-\frac{i}{\hbar }I_{S}\otimes H_{C}t^{\prime }}\left\vert \epsilon
_{S},t\right\rangle \left\langle \epsilon _{S},t\right\vert e^{\frac{i}{%
\hbar }I_{S}\otimes H_{C}t^{\prime }}=\frac{\chi (t+t^{\prime })}{\chi (t)}%
\left\vert \epsilon _{S},t+t^{\prime }\right\rangle \left\langle \epsilon
_{S},t+t^{\prime }\right\vert  \label{2a2}
\end{equation}%
\end{widetext}
which implies that the evolution operator $e^{-\frac{i}{\hbar }I_{S}\otimes
H_{C}t^{\prime }}$ acts on $\left\vert \epsilon _{S},t\right\rangle $ as%
\begin{equation}
e^{-\frac{i}{\hbar }I_{S}\otimes H_{C}t^{\prime }}\left\vert \epsilon
_{S},t\right\rangle =\sqrt{\frac{\chi (t+t^{\prime })}{\chi (t)}}\left\vert
\epsilon _{S},t+t^{\prime }\right\rangle  \label{2a3}
\end{equation}%
we can replace $t=0$ in last equation obtaining $e^{-\frac{i}{\hbar }%
I_{S}\otimes H_{C}t}\left\vert \epsilon _{S},0\right\rangle =\sqrt{\frac{%
\chi (t)}{\chi (0)}}\left\vert \epsilon _{S},t\right\rangle $. We can
proceed by computing $\frac{\partial }{\partial t}$ on the conjugate of the last equation and projecting on $\left\vert \psi \rangle \right\rangle $ and we
obtain a Schr\"{o}dinger-type equation
\begin{equation}
i\hbar \frac{\partial \xi }{\partial t}\left\langle \epsilon
_{S},t\right\vert \Psi \rangle \rangle +i\hbar \xi (t)\frac{\partial }{%
\partial t}\left\langle \epsilon _{S},t\right\vert \Psi \rangle \rangle
=\left\langle \epsilon _{S},t\right\vert H_{S}\otimes I_{C}\left\vert \Psi
\right\rangle \rangle  \label{2a4}
\end{equation}%
where $\xi (t)=\sqrt{\frac{\chi ^{\ast }(t)}{\chi ^{\ast }(0)}}$ and where
we have used that\ $I_{C}\otimes H_{S}=H-H_{C}\otimes I_{S}$. A
non-Hermitian term $i\frac{\partial \xi (t)}{\partial t}$ is added to the
Hamiltonian $H_{S}\otimes I_{C}$ and can be avoided by allowing
non-normalized states $\left\vert \psi ^{\prime }\right\rangle \rangle =\xi
(t)\left\vert \psi \right\rangle \rangle $ that satisfies%
\begin{equation}
i\hbar \frac{\partial }{\partial t}\left\langle \epsilon _{S},t\right\vert
\Psi ^{\prime }\rangle \rangle =\left\langle \epsilon _{S},t\right\vert \xi
^{-1}(t)H_{S}\otimes I_{C}\left\vert \Psi ^{\prime }\right\rangle \rangle
\label{2a5}
\end{equation}%
where the factor $\xi (t)$ is coupled to the Hamiltonian $H_{S}$. The extra coupling proportional to $\xi (t)$ originates
from the fact that the clock states $\{|t\rangle \}$ are not uniformly
normalized but weighted by $\chi (t)$. When projecting the global constraint
$H|\Psi \rangle \rangle =0$ onto $|\epsilon _{S},t\rangle $, the time
derivative acting on this non-normalized basis produces the extra term $%
i\hbar \,\partial _{t}\xi (t)$, which appears as a time-dependent coupling
in the effective system Hamiltonian. Redefining the conditional state as $%
|\Psi ^{\prime }\rangle \rangle =\xi (t)\,|\Psi \rangle \rangle $ with a
gauge transformation, the normalization factor is absorbed and the standard
Schr\"{o}dinger equation is restored governed solely by $\xi ^{-1}(t)H_{S}$.
The PaW mechanism finds the temporal evolution of a quantum system by its
dependence on some external clock time. The values of the observables of the
quantum system get correlated with the marking of the clock states with
definite clock times. The Page and Wootters mechanism fits naturally in the
Internal Quantum Reference Frames Programme (IQFR), where no reference
frames external to the whole universe exist, and ultimately we must define
the behaviour of subsystems of the universe relative to other subsystems of
the universe (\cite{giaco,castro,vanri,phili,aliahmad}).

\section{Time dilation in PaW mechanism}
\label{sec:td}

Typically physical systems are described relative to some fixed reference
frame which is understood as being external to the systems we aim to
describe. But in a closed system such as the whole universe, it is correct to
presume that there are no external reference frames. This implies that the
most we can do is to define the behaviour of subsystems of the universe
relative to other subsystems. Interestingly, systems composed of subsystems
allow one to decouple the center-of-mass motion from the internal dynamics
of the subsystems. This makes it possible to describe the system as two
non-interacting parts thereby enabling the application of the Page--Wootters
mechanism. Then, to introduce time dilation we must consider the Hamiltonian
of Eq.~(6) of Ref.~\cite{smith} [or Eq.~(9) of Ref.~\cite{chiba}]
\begin{equation}
H=\sqrt{P_{cm}^{2}c^{2}+M^{\prime 2}c^{4}}  \label{H1}
\end{equation}%
where $M^{\prime }=M+H_{r}/c^{2}$ is the mass of the composite system, $%
H_{r} $ is the internal energy and $P_{cm}$ is the center of mass momentum.
The internal energy contributes to the inertial mass, as dictated by mass-energy equivalence. The non-relativistic limit gives $H=\frac{P_{cm}^{2}%
}{2M^{\prime }}+H_{r}+O(P_{cm}^{4})$, which is identical to the modified
non-relativistic Hamiltonian obtained in Eq.~(1) of Ref.~\cite{paige} and where we
have neglected the rest energy. By using the identity $\frac{1}{x+y}=\frac{1
}{x}-\frac{y}{x(x+y)}$ the last equation can be written as
\begin{equation}
H=\frac{P_{cm}^{2}}{2M}+(I-\frac{P_{cm}^{2}}{2MM^{\prime }c^{2}})H_{r}
\label{8}
\end{equation}%
where the potential ambiguity with operator ordering is not a problem
because the internal Hamiltonian $H_{r}$ commutes with the center of mass
momentum $[P_{cm},H_{r}]=0$, since both operators belong to different Hilbert
spaces. This modification was introduced and used to study quantum
mechanical proper time \cite{zych}, and to resolve paradoxes in quantum
optics (\cite{sonn} and \cite{barn}). The factor $(I-\frac{P_{cm}^{2}}{%
2MM^{\prime }c^{2}})H_{r}$ can be written as $\left[ I-\frac{P_{cm}^{2}}{2M^{2}c^{2}}(1+\frac{H_{r}}{Mc^{2}})^{-1}\right] H_{r}$ and expanding $(1+\frac{%
H_{r}}{Mc^{2}})^{-1}$ in Taylor series and isolating the term linear in $%
H_{r}$, the Hamiltonian of Eq.~(\ref{8}) can be written as%
\begin{equation}
H=\frac{P_{cm}^{2}}{2M}\otimes \Omega (H_{r})+\gamma (P_{cm})\otimes H_{r}
\label{8a}
\end{equation}%
where $\Omega (H_{r})=\frac{H_{r}}{Mc^{2}}+(I+\frac{H_{r}}{Mc^{2}})^{-1}$
and $\gamma (P_{cm})=I-\frac{P_{cm}^{2}}{2M^{2}c^{2}}$. The factor $\gamma
(P_{cm})$ is, in principle, the quantum mechanical proper time
modification.\ By replacing $P_{cm}=Mv_{cm}$ the factor $\gamma $ can be
written as $\gamma =1-\frac{\beta ^{2}}{2}$ where $\beta =v_{cm}/c$. This
factor is the Taylor expansion in $\beta $ of the time dilation factor $%
\sqrt{1-\beta ^{2}}$.
Proceeding in a similar manner to the PaW mechanism,
Eq.~(\ref{8a}) is suitable to analyze how time dilation affects the center of
mass motion by considering a Hilbert space $\mathcal{H}=\mathcal{H}%
_{cm}\otimes \mathcal{H}_{r}$ where the internal quantum system $\mathcal{H}%
_{r}$ is the quantum clock system and $\mathcal{H}_{cm}$ is the center of
mass system. In contrast with Eq.~(\ref{1a}), the identity operators are
replaced by correction factors $\gamma (P_{cm})$ and $\Omega (H_{r})$ \cite{alex,rijavec}. We
can use the identity resolution Eq.~(\ref{6}) or Eq.~(\ref{0.8}) and it is not difficult to show that%
\begin{equation}
\begin{gathered}
e^{-\frac{i}{\hbar }\gamma (P_{cm})\otimes H_{r}t^{\prime }}I_{cm}\otimes Te^{\frac{i}{\hbar }\gamma (P_{cm})\otimes H_{r}t^{\prime }} = \\
I_{cm}\otimes T-t^{\prime }\gamma (P_{cm})\otimes I_{r}
\end{gathered}
\label{10}
\end{equation}
where we have used that $\left[ T,H_{r}\right] =i\hbar I_{r}$. Introducing
an identity operator in $\mathcal{H}_{cm}$ as $I_{cm}=\int dp_{cm}\left\vert
\mathbf{p}_{cm}\right\rangle \left\langle \mathbf{p}_{cm}\right\vert $ where
$H_{cm}\left\vert \mathbf{p}_{cm}\right\rangle =\epsilon _{cm}\left\vert
\mathbf{p}_{cm}\right\rangle $ and $\epsilon _{cm}=\frac{p_{cm}^{2}}{2M}$ is
the kinetic energy of the center of mass system, we can write\ $\gamma
(P_{cm})$ in the same basis as $\gamma (P_{cm})=\int d^{3}\mathbf{p}%
_{cm}\gamma (p_{cm})\left\vert \mathbf{p}_{cm}\right\rangle \left\langle
\mathbf{p}_{cm}\right\vert $, where $\gamma (p_{cm})=1-\frac{%
1}{2}(\frac{p_{cm}}{Mc})^{2}$ is the Lorentz factor. Following the same
procedure as in the last section, we obtain%
\begin{equation}
e^{-\frac{i}{\hbar }\gamma (P_{cm})\otimes H_{r}t^{\prime }}\left\vert
\mathbf{p}_{cm},t\right\rangle =\sqrt{\frac{\chi (t+\gamma t^{\prime })}{%
\chi (t)}}\left\vert \mathbf{p}_{cm},t+\gamma (p_{cm})t^{\prime
}\right\rangle  \label{11}
\end{equation}%
where we have used that $I_{r}=\int\limits_{{}}^{{}}dt\chi (t)\left\vert
t\right\rangle \left\langle t\right\vert $ is the identity operator in the
clock system. The Schr\"{o}dinger-type equation reads%
\begin{widetext}
\begin{equation}
i\hbar \xi (\gamma t)\frac{\partial }{\partial t}\left\langle \mathbf{p}%
_{cm},\gamma t\right\vert \Psi \rangle \rangle +i\hbar \frac{\partial \xi
(\gamma t)}{\partial t}\left\langle \mathbf{p}_{cm},\gamma t\right\vert \psi
\rangle \rangle =\left\langle \mathbf{p}_{cm},\gamma t\right\vert
H_{cm}\otimes \Omega (H_{r})\left\vert \psi \rangle \right\rangle
\label{11.1}
\end{equation}%
\end{widetext}
where $\xi (\gamma t)=\sqrt{\frac{\chi (\gamma t)}{\chi (0)}}$ and for the
unnormalized state $\left\vert \Psi ^{\prime }\right\rangle \rangle =\xi
(\gamma t)\left\vert \Psi \right\rangle \rangle $ the Schr\"{o}dinger-type
equation reduces to%
\begin{equation}
i\hbar \frac{\partial }{\partial t}\left\langle \mathbf{p}_{cm},\gamma
t\right\vert \Psi ^{\prime }\rangle \rangle =\left\langle \mathbf{p}%
_{cm},\gamma t\right\vert \xi ^{-1}(\gamma t)H_{cm}\otimes \Omega
(H_{r})\left\vert \Psi ^{\prime }\rangle \right\rangle  \label{12}
\end{equation}%
If we consider the limit $H_{r}/Mc^{2}\ll1$, then $\Omega (H_{r})=I_{r}$ and
we obtain Eq.~(\ref{2a5}). The gauge transformation $\xi ^{-1}(\gamma t)$
depends now on the time dilation $\gamma $ and $\Omega (H_{r})$ appears as a
normalization factor correction.
We note that the normalization factor acting on the center of mass depends
on the energy of the internal system, which represents a back-reaction of
the relative system on the center-of-mass degrees of freedom. Whereas
previous treatments of composite quantum clocks incorporate the internal
energy only through the linear proper-time coupling $\gamma (P_{cm})H_{r}$
\cite{zych,pikovski,smith,paige,grochowski}, here it also enters through the
normalization factor $\Omega (H_{r})$ acting on the center-of-mass sector.

For simplicity we can consider a simple example of how the normalization
factor affects the dynamical equation for the center of mass system. We can
consider the continuous cyclic quantum clock and we can assume that the
clock Hamiltonian is the internal energy of some quantum system. The
continuous cyclic quantum clock can be modeled by considering the quantum
states $\left\vert \phi \right\rangle =\frac{1}{\sqrt{2\pi }}%
\sum\limits_{n=-\infty }^{\infty }e^{in\phi }\left\vert n\right\rangle $
where $\left\vert n\right\rangle $ are eigenstates of $H_{r}=\omega L_{z}$
with eigenvalues $\hbar \omega n$. These states form a basis of the Hilbert
space and the identity can be written as $I=\sum\limits_{n=-\infty }^{\infty
}\left\vert n\right\rangle \left\langle n\right\vert =\int_{0}^{2\pi
}\left\vert \phi \right\rangle \left\langle \phi \right\vert d\phi $. We can
define a time operator $T$ as $T=\int_{0}^{2\pi }\phi \left\vert \phi
\right\rangle \left\langle \phi \right\vert d\phi $ and we have that $\left[
T,H_{r}\right] =i\hbar \omega I$. It is not difficult to show that
\begin{equation}
i\hbar \frac{\partial }{\partial t}\Psi _{cm}(\gamma \omega t)=\epsilon
_{cm}\int d\phi ^{\prime }\left\langle \gamma \omega t\right\vert \Omega
(H_{r})\left\vert \phi ^{\prime }\right\rangle \Psi _{cm}(\phi ^{\prime })
\label{k1}
\end{equation}%
which is a non-local integral Schr\"{o}dinger equation. This is the main effect
of the $\Omega (H_{r})$, since the kernel $\left\langle \gamma \omega
t\right\vert \Omega (H_{r})\left\vert \phi ^{\prime }\right\rangle $
prevents the wave function $\Psi _{cm}(\phi ^{\prime })$ from being evaluated at $%
\gamma \omega t$ on the right-hand side of the last equation. Time non-local Schr\"{o}dinger equations of this type have appeared before in
the Page--Wootters framework when an explicit interaction between the clock
and the system is present \cite{alex,mendes}, and a term quadratic in the
Hamiltonian has been obtained from a gravitationally induced clock--system
coupling with a non-ideal clock \cite{mendes}. Here, by contrast, the
non-locality and the quadratic energy dependence originate from the
normalization factor $\Omega(H_{r})$ that comes from the mass-energy back-reaction of the
free composite particle. This kernel can be written
as
\begin{equation}
\left\langle \omega t\right\vert \Omega (H_{r})\left\vert \phi ^{\prime
}\right\rangle =\frac{1}{2\pi }\sum\limits_{n=-\infty }^{\infty
}e^{in(\gamma \omega t-\phi )}(\eta n+\frac{1}{1+\eta n})  \label{k2}
\end{equation}%
where $\eta =\frac{\hbar \omega }{Mc^{2}}$. Splitting the sum into the term
$n=0$ and the contributions $n\geq 1$ and $n\leq -1$, writing $n=-m$ in the
latter and using $(1+\eta n)(1-\eta n)=1-\eta ^{2}n^{2}$, the kernel takes the
real trigonometric form
\begin{widetext}
\begin{equation}
\left\langle \omega t\right\vert \Omega (H_{r})\left\vert \phi ^{\prime
}\right\rangle =\frac{1}{2\pi }\left[ 1+2i\eta \sum_{n=1}^{\infty }n\sin
(n\Delta )+2\sum_{n=1}^{\infty }\frac{\cos (n\Delta )-i\eta \,n\sin (n\Delta )%
}{1-\eta ^{2}n^{2}}\right] ,  \label{k2b}
\end{equation}%
\end{widetext}
with $\Delta =\gamma \omega t-\phi $. The first sum is the local part: since
$\sum_{n=1}^{\infty }n\sin (n\Delta )=-\pi \delta ^{\prime }(\Delta )$ (mod $%
2\pi $), it reduces to $-i\eta \,\delta ^{\prime }(\Delta )$, while the
convergent sums weighted by $(1-\eta ^{2}n^{2})^{-1}$ encode the smooth
non-local contribution. In the ideal-clock limit $\eta \rightarrow 0$ the real
part collapses to $\frac{1}{2\pi }[1+2\sum_{n\geq 1}\cos (n\Delta )]=\delta
(\Delta )$ and the kernel becomes local.
There are different approaches to solve Eq.~(\ref{k1}) or at least
to describe the non-locality in the time variable, which in this example is
the angle $\phi $. For instance, solving the infinite sums we obtain
\begin{widetext}
\begin{equation}
\left\langle \omega t\right\vert \Omega (H_{r})\left\vert \phi ^{\prime
}\right\rangle =\frac{1}{2\pi \eta }\left[ \Phi (e^{i\Delta },1,\eta
^{-1})-e^{i\Delta }\Phi (e^{-i\Delta },1,1-\eta ^{-1})\right] -i\eta \delta
^{\prime }(\Delta )  \label{k3}
\end{equation}%
\end{widetext}
where $\Phi (z,s,a)$ is the Lerch function and $\Delta =\phi ^{\prime
}-\omega t$. The kernel is defined for all possible values of $\Delta $ mod $%
2\pi $ which implies that non-locality extends over the whole period $T=\frac{%
2\pi }{\omega }$. Nevertheless, the scale $\tau =\frac{\hbar }{Mc^{2}}=\frac{%
\eta }{\omega }$ controls the effective non-locality in time, since in the $%
0<\Delta <2\pi $ branch, the kernel behaves as $\left\langle \omega
t\right\vert \Omega (H_{r})\left\vert \phi ^{\prime }\right\rangle \sim e^{-%
\frac{i\Delta }{\eta }}$. This implies that the kernel oscillates with
characteristic time scale $\tau $, so contributions from intervals much
larger than $\tau $ tend to cancel by phase interference. Therefore the
effective non-locality is of order $\tau $ and it is interesting to note that this effective temporal non-locality is the time analogue of the Compton
wavelength, that sets the spatial non-locality scale of a particle in relativistic quantum mechanics. This may
indicate that particle non-locality is more naturally understood as a
covariant phenomenon, with spatial and temporal nonlocal aspects emerging
within a unified description.

\subsection{Examples of clock-systems}

Relativistic point-like systems with internal dynamics can be viewed as
ideal clocks, since their internal degrees of freedom evolve according to
the proper time along the system's worldline and therefore register the
elapsing proper time. In what follows, we work in the regime in which the
total energy $H$ of Eq.~(\ref{1a}) can
be approximated as the total energy of Eq.~(\ref{8a}), as discussed in the previous section, and is a generalization of the approximation used in Ref.~\cite{paige} [see Eq.~(1)]. This should not be understood as a
purely non-relativistic approximation in which all relativistic effects are
discarded. Rather, we consider a regime in which the center-of-mass motion
can be treated in a Newtonian sense, while the contribution of the internal
energy to the rest mass is still retained through mass-energy equivalence.
Technically, there are two distinct small parameters, $\frac{H_{r}}{Mc^{2}}$
and $\frac{P_{cm}^{2}}{M^{2}c^{2}}$ associated with internal and motional
relativistic corrections, respectively. The approximation therefore amounts
to neglecting higher-order motional corrections while keeping the
internal-energy contribution to the effective mass. For simplicity, we shall
consider two main examples: two quantum particles coupled by a spring, and
the hydrogen atom. Both systems satisfy the condition required to implement
the Page--Wootters mechanism, since in the non-relativistic regime the
center-of-mass degrees of freedom decouple from the relative motion,
allowing the formalism of the previous section to be applied.

\subsubsection{Two quantum particles coupled with a spring}

In this case the Hamiltonian reads
\begin{equation}
H=\frac{p_{1}^{2}}{2m_{1}}+\frac{p_{2}^{2}}{2m_{2}}+\frac{1}{2}\omega
^{2}(x_{1}-x_{2})^{2}  \label{0.1}
\end{equation}%
where we are considering only one spatial dimension. Introducing the
position, momentum and mass associated with the center of mass system as $%
x_{cm}=(m_{1}x_{1}+m_{2}x_{2})/M$, $p_{cm}=p_{1}+p_{2}$ where $M=m_{1}+m_{2}$
and the position, momentum and the mass of the relative system $%
x_{r}=x_{1}-x_{2}$, $p_{r}=(m_{2}p_{1}-m_{1}p_{2})/M$ and $\mu _{r}=\frac{%
m_{1}m_{2}}{M}$ the effective mass, the Hamiltonian can be written as
\begin{equation}
H=\frac{p_{cm}^{2}}{2M}+\frac{p_{r}^{2}}{2\mu _{r}}+\frac{1}{2}\omega
^{2}x_{r}^{2}  \label{0.2}
\end{equation}%
which can be written as $H=H_{cm}\otimes I_{r}+I_{cm}\otimes H_{r}$, where $%
H_{cm}=\frac{p_{cm}^{2}}{2M}$ and $H_{r}=\frac{p_{r}^{2}}{2\mu _{r}}+\frac{1%
}{2}\omega ^{2}x_{r}^{2}$ and are the well-known Hamiltonians of a free
particle for the center of mass system $H_{cm}$ and the Hamiltonian of a
quantum oscillator~$H_{r}$ for the relative system. At this point is
important to clarify that Hamiltonian of Eq.~(\ref{0.1}) is the limit $\frac{%
H_{r}}{Mc^{2}}\rightarrow 0$, $\frac{P_{cm}}{Mc}\rightarrow 0$ of the
relativistic energy $\sqrt{P_{cm}^{2}c^{2}+M^{\prime 2}c^{4}}$, where $%
M^{\prime }=M+\frac{H_{r}}{c^{2}}$ and where only the rest energy is
neglected in the final result. The last Hamiltonian is of the form required for
the PaW mechanism, where no interaction is between the subsystems. We can
define the resolution of unity in $H_{r}$ using coherent states of the
quantum harmonic oscillator%
\begin{equation}
I_{r}=\frac{1}{\pi }\int_{0}^{\infty }\int_{0}^{2\pi }\rho dpd\phi
\left\vert \rho ,\phi \right\rangle \left\langle \rho ,\phi \right\vert
\label{0.4}
\end{equation}%
where $\left\vert \rho ,\phi \right\rangle =e^{-\frac{\rho ^{2}}{2}%
}\sum\limits_{n=0}^{\infty }\frac{\rho ^{n}e^{in\phi }}{\sqrt{n!}}\left\vert
n\right\rangle $ are the coherent states defined in the basis $\left\vert
n\right\rangle $ that diagonalizes $H_{r}$, that is $H_{r}\left\vert
n\right\rangle =\epsilon _{n}\left\vert n\right\rangle $, where $\epsilon
_{n}=\hbar \omega (n+\frac{1}{2})$. These coherent states are temporally stable because they transform covariantly under the group action $e^{-\frac{i}{\hbar }H_{r}t}\left\vert \rho ,\phi \right\rangle =\left\vert
\rho ,\phi -\omega t\right\rangle $ so these coherent states can be used as quantum clock states. In \cite%
{foti} generalized coherent states are used as the clock states and the
phase [see Eq.~(5) of Ref.~\cite{foti}] plays the role of classical and quantum
time, that is, the phase of the coherent states follows an ellipse in phase space and is experimentally observable. A closely related construction is the parametric representation of composite quantum systems, in which one subsystem is described through generalized
coherent states of the complementary subsystem, weighted by a coherent-state
amplitude, providing an exact description of the composite dynamics
\cite{calvani}. As $H_{r}$ acts as a time-translation operator, the conjugate operator $T$ obeying the relation $\left[
T,H_{r}\right] =i\hbar I_{r}$ can be obtained and in Appendix A it is shown
that it can be written as
\begin{equation}
T=-\frac{1}{\pi \omega }\int \rho d\rho \int d\phi \phi \left\vert \rho
,\phi \right\rangle \left\langle \rho ,\phi \right\vert  \label{0.6}
\end{equation}%
As it was discussed above, starting from the relativistic energy, the most
appropriate splitting of the Hamiltonian is the one obtained in Eq.~(\ref{8a}%
). Using Eq.~(\ref{10}) we have%
\begin{equation}
e^{-\frac{i}{\hbar }\gamma (P_{cm})\otimes H_{r}t}\left\vert \mathbf{p}%
_{cm}\right\rangle \otimes \left\vert \rho ,\phi \right\rangle =\left\vert
\mathbf{p}_{cm}\right\rangle \otimes \left\vert \rho ,\phi -\gamma
(p_{cm})\omega t\right\rangle  \label{0.8}
\end{equation}%
Conjugating the last result and projecting over the quantum state $\left\vert
\Psi \rangle \right\rangle $ of the global system we obtain that%
\begin{widetext}
\begin{equation}
i\hbar \frac{\partial }{\partial t}\left\langle \mathbf{p}_{cm}\right\vert
\otimes \left\langle \rho ,-\gamma (p_{cm})\omega t\right\vert \Psi
\rangle \rangle =\left\langle \mathbf{p}_{cm}\right\vert \otimes
\left\langle \rho ,-\gamma (p_{cm})\omega t\right\vert H_{cm}\otimes \Omega
(H_{r})\left\vert \Psi \right\rangle \rangle  \label{0.9}
\end{equation}%
Finally, expanding the quantum global state $\left\vert \Psi \right\rangle
\rangle $ in the basis $\left\vert \mathbf{p}_{cm}\right\rangle \otimes
\left\vert \rho ,\phi \right\rangle $ and using that $\frac{\partial }{%
\partial t}=-\omega \gamma \frac{\partial }{\partial \phi }$, the last equation can be recast in Schr\"{o}dinger form%

\begin{equation}
-i\hbar \gamma \omega \frac{\partial }{\partial \phi }\Psi _{cm}(\rho ,\phi
)=\frac{p_{cm}^{2}}{2M}\int \rho ^{\prime }d\rho ^{\prime }d\phi ^{\prime
}\left\langle \rho ,\phi \right\vert \Omega (H_{r})\left\vert \rho ^{\prime
},\phi ^{\prime }\right\rangle \Psi _{cm}(\rho ^{\prime },\phi ^{\prime })
\label{1.1}
\end{equation}%
where $\Psi _{cm}(\rho ,\phi )=\left\langle \rho ,\phi \right\vert
\left\langle \mathbf{p}_{cm}\right\vert \Psi \rangle \rangle $. In this case, the kernel $\left\langle \rho ,\phi \right\vert \Omega
(H_{r})\left\vert \rho ^{\prime },\phi ^{\prime }\right\rangle $ is more
involved due to the non-orthogonality of the coherent states. Using the
definition, the kernel reads
\begin{gather}
\left\langle \rho ,\phi \right\vert \Omega (H_{r})\left\vert \rho ^{\prime
},\phi ^{\prime }\right\rangle =e^{-\frac{1}{2}(\rho ^{2}+\rho ^{\prime
2})}\sum\limits_{n=0}^{\infty }\frac{(\rho ^{\prime }\rho e^{i(\phi ^{\prime
}-\phi )})^{n}}{n!}\left[ \eta (n+\frac{1}{2})+\frac{1}{1+\eta (n+\frac{1}{2}%
)}\right]   \label{1.2} \\
=e^{-\frac{1}{2}(\rho ^{2}+\rho ^{\prime 2})}\left[ \frac{\eta }{2}(1+2\rho
^{\prime }\rho e^{i(\phi ^{\prime }-\phi )})e^{\rho ^{\prime }\rho e^{i(\phi
^{\prime }-\phi )}}+\frac{1}{\eta }(-\rho ^{\prime }\rho e^{i(\phi ^{\prime
}-\phi )})^{-\frac{\eta +2}{2\eta }}\Gamma (\frac{1}{2}+\frac{1}{\eta }%
,0,-\rho ^{\prime }\rho e^{i(\phi ^{\prime }-\phi )})\right]   \notag
\end{gather}%
\end{widetext}
where $\Gamma (x,y,z)$ is the generalized incomplete $\Gamma $ function and $%
\eta _{HO}=\frac{\hbar \omega }{Mc^{2}}$.

\subsubsection{Hydrogen atom}

An identical procedure can be applied to the hydrogen atom, which can
be separated as $H=H_{CM}\otimes I_{r}+I_{CM}\otimes H_{r}$, where $H_{cm}=%
\frac{P_{cm}^{2}}{2M}$ is the center of mass Hamiltonian which is purely kinetic and $H_{r}=\frac{P_{r}^{2}}{2\mu }+V(r)$ and where $M$ and $\mu$ are the total and effective mass of the center of mass system and the relative system respectively. This Hamiltonian $H$ is again of the form required for the PaW mechanism, where no interaction is between the subsystems. The $I_{cm}$ and $I_{r}$ are the identity operators in the respective Hilbert spaces $H_{cm}$ and $H_{r}$. The total Hilbert space can be written as $\mathcal{H}=\mathcal{H}%
_{cm}\otimes \mathcal{H}_{R}$.
Following the PaW mechanism, let us consider the relative state of motion
between the proton and electron as the quantum system that is used as a
quantum clock. Considering that the proton and electron are in a bound
state, then we have at our disposal the identity resolution%
\begin{equation}
I_{r}=\underset{n=0}{\overset{\infty }{\sum }}\underset{l=0}{\overset{n}{%
\sum }}\underset{m=-l}{\overset{l}{\sum }}\left\vert n+1,l,m\right\rangle
\left\langle n+1,l,m\right\vert   \label{3}
\end{equation}%
where $\left\vert n+1,l,m\right\rangle $ are the eigenstates of $H_{r}$ that
satisfies $H_{r}\left\vert n+1,l,m\right\rangle =\epsilon _{n}\left\vert
n+1,l,m\right\rangle $, where $\epsilon _{n}=\frac{\kappa ^{2}}{(n+1)^{2}}$,
$\kappa ^{2}=\frac{\mu e^{4}}{2(4\pi \epsilon _{0})^{2}\hbar ^{2}}$ and $%
\left\langle r,\theta ,\phi \right\vert n+1,l,m\rangle
=R_{n+1}(r)Y_{lm}(\theta ,\phi )$ are the well-known hydrogen atom states in
coordinate representation. In turn, we have at our disposal hydrogen atom
coherent states $\left\vert s,\varsigma \right\rangle $ that can be written in
terms of the $\left\vert n+1,l,m\right\rangle $ basis as (see \cite{curado})%
\begin{equation}
\left\vert s^{\prime },\varsigma ^{\prime }\right\rangle =N(s^{\prime })%
\underset{n=1}{\overset{\infty }{\sum }}(s^{\prime })^{n-1}\frac{\sqrt{2n}%
e^{i\frac{\varsigma ^{\prime }}{n^{2}}}n}{\sqrt{n+1}}\left\vert
n\right\rangle   \label{4}
\end{equation}%
where $\left\vert n\right\rangle =\frac{1}{n}\underset{l=0}{\overset{n-1}{%
\sum }}\underset{m=-l}{\overset{l}{\sum }}\left\vert n,l,m\right\rangle $, $%
s^{\prime }=\frac{s}{\kappa }\in \left[ 0,1\right] $ and $\varsigma ^{\prime
}=\varsigma \kappa ^{2}$. The normalization factor can be written as%

\begin{equation}
\begin{gathered}
N^{2}(s^{\prime})= \\
\frac{
(s^{\prime})^{4}
\left(1-(s^{\prime})^{2}\right)^{3}
}{
2\left[
3(s^{\prime})^{6}
-2(s^{\prime})^{4}
+(s^{\prime})^{2}
\right]
+
2\left(1-(s^{\prime})^{2}\right)^{3}
\ln\left(1-(s^{\prime})^{2}\right)
}
\end{gathered}
\label{5}
\end{equation}
and the resolution of identity can be written as%
\begin{equation}
I_{r}=\underset{L\rightarrow \infty }{\lim }\frac{1}{2\pi }%
\int\limits_{-L}^{L}d\varsigma ^{\prime }\int\limits_{0}^{1}ds^{\prime }\rho
(s^{\prime })\left\vert s^{\prime },\varsigma ^{\prime }\right\rangle
\left\langle s^{\prime },\varsigma ^{\prime }\right\vert =\underset{n=1}{%
\overset{\infty }{\sum }}\left\vert n\right\rangle \left\langle n\right\vert
\label{6}
\end{equation}%
where $\rho (s^{\prime })=\frac{2s^{\prime }\ln s^{\prime }}{N^{2}(s^{\prime
})}[\ln s^{\prime }-1]$ is the weight function. The group action $e^{-\frac{i%
}{\hbar }H_{r}{}^{\prime }}$ with $H_{r}$ as the generator acting on $%
\left\vert s,\varsigma \right\rangle $ gives $e^{-\frac{i}{\hbar }%
H_{C}^{\prime }\varsigma ^{\prime }}=\left\vert s,\varsigma +\varsigma
^{\prime }\right\rangle $ as it is expected for the PaW mechanism. Since $%
\varsigma $ acts as a time parameter, the kernel can be written as
\begin{widetext}
\begin{equation}
\left\langle s,\varsigma \right\vert \Omega (H_{r})\left\vert s,\varsigma ^{\prime
}\right\rangle =2N^{2}(s)\underset{n=1}{\overset{\infty }{\sum }}\frac{%
n^{3}s^{2(n-1)}e^{\frac{i}{n^{2}}(\varsigma -\varsigma ^{\prime })}}{(n+1)}%
\left[ \frac{\eta _{HA}}{n^{2}}+\left(1+\frac{\eta _{HA}}{n^{2}}\right)^{-1}\right]
\label{eq:kernelH}
\end{equation}%
\end{widetext}
where for simplicity we replace $s^{\prime }=s$ since it is a parameter
analogous to $\rho $ in the coherent states of the harmonic oscillator. The
analogous dimensionless factor is $\eta _{HA}=\frac{\kappa ^{2}}{Mc^{2}}\sim
\frac{m_{e}}{2m_{p}}\alpha ^{2}$, where $\alpha $ is the fine structure
constant.
The kernel behavior of eq.(\ref{eq:kernelH}) is qualitatively different from that of the cyclic and harmonic-oscillator clocks. In these quantum clocks, the spectrum is equidistant,
$E_{n}\propto n$, so increasing the truncation $N$ in the sum of eq.(\ref{eq:kernelH}) adds ever more rapidly
oscillating phases and thus widens the spectral bandwidth available to
localize the kernel around $\varsigma=\varsigma^{\prime}$. For the hydrogen
clock the situation is reversed: the bound-state energies tends to $0$ as $n\to\infty$, so the phase $e^{i(\varsigma-\varsigma^{\prime})/n^{2}}$ of the $n$-th term oscillates ever
more slowly. Raising $N$ therefore supplies progressively lower frequencies
rather than higher ones, the available bandwidth is bounded from above by the
$n=1$ frequency, and no number of additional levels can sharpen the kernel since its width in $\varsigma-\varsigma^{\prime}$ does not shrink as $N$ grows.
The convergence in $N$ is further accelerated by the coherent-state weights.
Isolating the $\Omega$-induced part of the kernel, the $n$-th coefficient is $a_{n}^{H}\propto\frac{n^{3}}{n+1}\,s^{2(n-1)}
\left[\Omega\!\left(\frac{\eta_{HA}}{n^{2}}\right)-1\right]$. Since $\eta_{HA}/n^{2}\ll1$ and $\Omega(x)-1=x^{2}/(1+x)\simeq x^{2}$, this
reduces to $a_{n}^{H}\sim\frac{\eta_{HA}^{2}}{n(n+1)}\,s^{2(n-1)}$ so successive terms are suppressed geometrically by $s^{2}$. The sum is thus
dominated by its first few modes and is effectively converged well before
$n\sim10$; the factor $s^{2(n-1)}$ acts as a natural spectral regulator. The hydrogen kernel then converges rapidly in the truncation $N$, but it converges to a broad profile in
$\varsigma-\varsigma^{\prime}$, that is, the hydrogen clock localizes clock time less sharply than the equidistant clocks.

\subsection{Momentum superposition, internal clock, and interferometric visibility}

We can now apply the previous formalism to an interferometric situation in which
the center of mass of a composite quantum system is prepared in a superposition
of two different momenta (\cite{zych}, \cite{marga}, \cite{loriani}). The physical procedure is the following: a composite particle, such as a hydrogen atom or a two-particle bound system, possesses internal relative degrees of freedom that define
a quantum clock. A beam splitter acting on the center of mass prepares a
superposition of two momentum branches. Experimentally, a momentum superposition can be generated by coherently coupling different center-of-mass momentum states. A beam-splitting pulse, implemented through diffraction from a spatially periodic potential or field, transfers a controlled momentum while preserving quantum coherence. When the pulse area is appropriately chosen, the initial wave packet is transformed into a coherent superposition of two components with distinct momenta \cite{Cronin2009,kovachy}.
Since the two momenta are defined with
respect to the laboratory frame, the discussion of visibility must also be
formulated in the laboratory frame. Then, let \(t\) denote the coordinate time of the laboratory. The center-of-mass momenta \(p_1\) and \(p_2\) are laboratory-frame momenta. Each momentum branch defines a different proper time for the composite system. If \(\tau_i\) denotes
the proper time accumulated along the branch with momentum \(p_i\), then $d\tau_i=\gamma_i\,dt$ and $\gamma_i\equiv\gamma(p_i)$, where \(\gamma_i\) is the time-dilation factor associated with that branch. In the nonrelativistic regime one may write $\gamma_i\simeq 1-\frac{p_i^2}{2M^2c^2}$.
In the simplest internal clock discussed in the previous section, the clock variable is a cyclic phase \(\phi\). In the
proper frame of the composite system, the phase of the clock evolves according
to $d\phi=\omega\,d\tau$. Therefore, in the laboratory frame, the phase associated with the branch \(p_i\) satisfies $d\phi_i=\omega\,d\tau_i=\omega\gamma_i\,dt$. If the initial clock reading is chosen to be \(\phi(0)=0\), then, for a branch
with constant momentum \(p_i\), we can write $\phi_i(t)=\omega\gamma_i t$. Thus a laboratory observer sees different CM momentum branches correlated
with different internal clock readings. This correlation is the origin of
which-branch information and therefore of the possible loss of interferometric
visibility.
For simplicity, we take the internal clock Hamiltonian of the cyclic clock and we can write
a general initial clock state as
\begin{equation}
|\chi_0\rangle=
\sum_{n=-\infty}^{\infty}f_n|n\rangle,
\label{eq:clock-initial-n}
\end{equation}
where $\sum_n |f_n|^2=1$. In the phase representation we have $F_0(\phi)=\langle \phi|\chi_0\rangle
=
\frac{1}{\sqrt{2\pi}}
\sum_{n=-\infty}^{\infty}f_n e^{-in\phi}$.
The statement that the clock is initially set at \(\phi(0)=0\) means that
\(F_0(\phi)\) is localized around \(\phi=0\). The ideal limit would be a
periodic delta distribution, but a normalizable phase packet is physically
more appropriate \cite{woods2019}.
We now consider the interferometric preparation. An atomic beam splitter,
implemented for instance by a coherent Bragg or Raman pulse, prepares the CM
state in a superposition of two laboratory momenta $|\mathbf{p}_0\rangle
\longrightarrow
\frac{1}{\sqrt{2}}
\left(|\mathbf{p}_1\rangle+e^{i\varphi_{\rm BS}}|\mathbf{p}_2\rangle
\right)$. The phase \(\varphi_{\rm BS}\) is controlled by the laser fields and may be
absorbed into the definition of one branch. We therefore take the state after
the first beam splitter to be
\begin{equation}
|\Psi(0)\rangle=
\frac{1}{\sqrt{2}}
\left(
|\mathbf{p}_1\rangle+|\mathbf{p}_2\rangle
\right)
\otimes|\chi_0\rangle .
\label{eq:initial-interferometer}
\end{equation}
Projecting Eq.~\eqref{8a} onto the branch \(|\mathbf{p}_i\rangle\), the
internal clock evolves with the conditional Hamiltonian $H_i^{\rm int}
=\gamma_i H_r+\epsilon_i\Omega(H_r)$, where the first term shifts the phase of the clock and the second term is the correction coming from the normalization factor \(\Omega(H_r)\). Thus, after a laboratory time
\(T\), the state of the clock correlated with the branch \(p_i\) in the energy basis of the cyclic clock is
\begin{equation}
|\chi_i(T)\rangle
=
\sum_{n=-\infty}^{\infty}
f_n
\exp\left[
-\frac{iT}{\hbar}
\left(
\gamma_i\hbar\omega n+\epsilon_i\Omega_n
\right)
\right]
|n\rangle,
\label{eq:clock-branch-n}
\end{equation}
where $\Omega_n=\Omega(\hbar\omega n)$. In the phase representation, Eq.~\eqref{eq:clock-branch-n} gives
\begin{widetext}
\begin{equation}
F_i(\phi,T)
=
\langle\phi|\chi_i(T)\rangle
=
\frac{1}{\sqrt{2\pi}}
\sum_{n=-\infty}^{\infty}
f_n
\exp\left[
-\frac{iT}{\hbar}
\left(
\gamma_i\hbar\omega n+\epsilon_i\Omega_n
\right)
\right]
e^{-in\phi}.
\label{eq:clock-branch-phase}
\end{equation}
\end{widetext}
If the normalization correction is neglected, \(\Omega_n\simeq1\), this
reduces to $F_i(\phi,T)
\simeq
e^{-i\epsilon_iT/\hbar}
F_0(\phi+\omega\gamma_i T)$.
We now recombine the two CM branches. An ideal final atomic beam splitter
acts as
$
|p_1\rangle
\longrightarrow
\frac{1}{\sqrt{2}}
\left(
|D_+\rangle+|D_-\rangle
\right)$ and $|p_2\rangle
\longrightarrow
\frac{1}{\sqrt{2}}
\left(
|D_+\rangle-|D_-\rangle
\right)$
where \(|D_+\rangle\) and \(|D_-\rangle\) denote the two output ports. Before
the final beam splitter, the state may be written as
\begin{equation}
|\Psi(T)\rangle
=
\frac{1}{\sqrt{2}}
\left(
|p_1\rangle\otimes|\chi_1(T)\rangle
+
e^{i\varphi_{\rm ext}}
|p_2\rangle\otimes|\chi_2(T)\rangle
\right),
\label{eq:pre-final-state}
\end{equation}
where \(\varphi_{\rm ext}\) contains any controllable external interferometric
phase, such as laser phases or propagation phases. After the final beam
splitter,
\begin{align}
|\Psi_{\rm out}(T)\rangle
&=
\frac{1}{2}
|D_+\rangle
\otimes
\left(
|\chi_1(T)\rangle
+
e^{i\varphi_{\rm ext}}|\chi_2(T)\rangle
\right)
\nonumber\\
&\quad+
\frac{1}{2}
|D_-\rangle
\otimes
\left(
|\chi_1(T)\rangle
-
e^{i\varphi_{\rm ext}}|\chi_2(T)\rangle
\right).
\label{eq:out-state}
\end{align}
Therefore the output probabilities are
\begin{equation}
P_\pm(T)
=
\frac{1}{2}
\left[
1
\pm
\mathrm{Re}
\left(
e^{i\varphi_{\rm ext}}\Phi(T)
\right)
\right],
\label{eq:output-probs}
\end{equation}
where $\Phi(T)=\langle \chi_2(T)|\chi_1(T)\rangle$. The interferometric visibility is $V(T)=|\Phi(T)|$.
Thus the loss of visibility is determined by the distinguishability of the
two internal clock states correlated with the two CM momentum branches.
Using Eq.~\eqref{eq:clock-branch-n}, we obtain
\begin{equation}
\Phi(T)
=
\sum_{n=-\infty}^{\infty}
|f_n|^2
\exp\left[
-\frac{iT}{\hbar}
\left(
\Delta\gamma\,\hbar\omega n
+
\Delta\epsilon\,\Omega_n
\right)
\right]
\label{eq:Phi-exact}
\end{equation}
with
\begin{equation}
\Delta\gamma=\gamma_1-\gamma_2,
\qquad
\Delta\epsilon=\epsilon_1-\epsilon_2.
\label{eq:Delta-defs}
\end{equation}
Equivalently, in the phase representation,
\begin{equation}
\Phi(T)
=
\int_0^{2\pi}d\phi\,
F_2^*(\phi,T)F_1(\phi,T).
\label{eq:Phi-phase}
\end{equation}
In the standard limit,
\(\Omega_n\simeq1\), Eq.~\eqref{eq:Phi-exact} becomes
\begin{equation}
\Phi(T)
\simeq
e^{-i\Delta\epsilon T/\hbar}
\sum_n
|f_n|^2 e^{-i\Delta\gamma\omega nT}.
\label{eq:Phi-standard}
\end{equation}
and the prefactor \(e^{-i\Delta\epsilon T/\hbar}\) is a global phase and does not
affect the visibility. Introducing the proper-time difference $\Delta\tau(T)=\Delta\gamma\,T$
we recover $V(T)
=
\left|
\langle \chi_0|
e^{-iH_r\Delta\tau(T)/\hbar}
|\chi_0\rangle
\right|$. This is the usual visibility formula for time-dilation-induced loss of
interference. The internal clock carries which-branch information because the
two CM momentum branches correspond to different proper times.

Keeping the correction \(\Omega(H_r)\) and using Eq.~\eqref{eq:Phi-exact} and removing the irrelevant global phase we have
\begin{equation}
V(T)
\simeq
\left|
\sum_n |f_n|^2
\exp\left[
-\frac{iT}{\hbar}
\left(
\Delta\gamma E_n
+
\frac{\Delta\epsilon}{M^2c^4}E_n^2
\right)
\right]
\right|
\label{eq:visibility-corrected}
\end{equation}
The first term in the phase is the standard time-dilation contribution. The
second term is the leading correction induced by the normalization factor
\(\Omega(H_r)\). It is quadratic in the internal energy and therefore depends
on the absolute distribution of clock energies, not only on the energy gaps.

\subsubsection{Explicit clock models and experimental estimates}

\begin{figure}[h]
\includegraphics[width=\columnwidth]{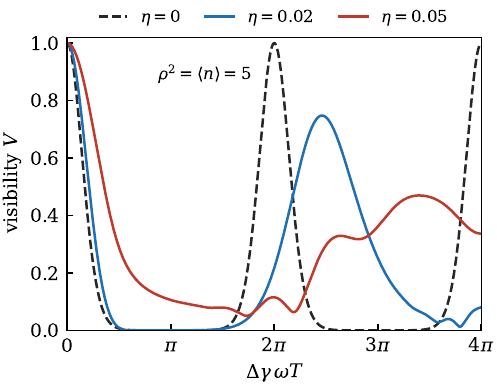}
\caption{Interferometric visibility $V$ of a coherent-state oscillator clock (mean excitation $\rho^{2}=\langle n\rangle=5$) versus the dimensionless time $\Delta\gamma\,\omega T$. The dashed curve ($\eta=0$) is the standard time-dilation result, which is periodic and fully revives at $\Delta\gamma\,\omega T=2\pi$. The solid curves include the normalization factor $\Omega(H_{r})$, whose quadratic-in-energy term (Eq.~(\ref{eq:visibility-corrected})) suppresses and shifts the revivals. For clarity, $\eta=\hbar\omega/Mc^{2}$ is set to illustrative values far larger than any atomic-clock value ($\eta\sim10^{-8}$), for which the two curves would be indistinguishable on this scale; the enhanced regime corresponds to the large coherent-internal-energy scenario discussed in the text.}
\label{fig:visibility}
\end{figure}

The general expression obtained above can now be evaluated for different
internal clocks. The only ingredients needed are the spectrum of \(H_r\) and
the energy distribution of the initial clock state.
For the harmonic oscillator clock, considering that the  internal state is initially prepared in a coherent state,
using the low-energy expansion of \(\Omega(E)\), one obtains
\begin{equation}
\Phi_{\rm osc}(t)
\simeq
e^{-i\Delta\epsilon t/\hbar}
\left\langle
\alpha_0
\right|
\exp\left[
-\frac{i t}{\hbar}
\left(
\Delta\gamma H_r
+
\frac{\Delta\epsilon}{M^2c^4}H_r^2
\right)
\right]
\left|
\alpha_0
\right\rangle
\end{equation}
Up to the global phase, $\Phi_{\rm osc}(t)$ is the characteristic function of
the operator $K=\Delta\gamma H_{r}+(\Delta\epsilon/M^{2}c^{4})H_{r}^{2}$
evaluated in the coherent state $|\alpha_0\rangle$, so the visibility
$V=|\Phi_{\rm osc}(t)|$ is controlled by the even cumulants of $K$ alone:
odd cumulants contribute only pure phases and cancel in the modulus, and to
leading order in $t$ the decay is fixed by $\operatorname{Var}_{\alpha_0}(K)$.
If the \(\Omega(H_r)\) correction is neglected, the coherent state undergoes
only a rigid phase-space rotation in each momentum branch. In that case,
using $\operatorname{Var}_{\alpha_0}(H_r)=
(\hbar\omega)^2\rho^2$ one obtains the standard Gaussian visibility decay $V_0(t)=
\exp\left[
-\frac{1}{2}
\left(
\rho\,\Delta\gamma\,\omega t
\right)^2
\right]$. The leading correction induced by \(\Omega(H_r)\) can be estimated from the
same short-time cumulant formula. Expanding $\operatorname{Var}_{\alpha_0}(K)$,
besides the standard term $(\Delta\gamma)^{2}\operatorname{Var}_{\alpha_0}(H_r)$, a cross term
$2\,\Delta\gamma\,(\Delta\epsilon/M^{2}c^{4})\operatorname{Cov}_{\alpha_0}(H_r,H_r^2)$ appears,
which is the leading effect of $\Omega$. Then the dominant correction comes from the correlation between the linear and quadratic phases rather than
from the dispersion of $H_r^2$ itself. For a coherent oscillator state one also has $\operatorname{Cov}_{\alpha_0}(H_r,H_r^2)
=
2(\hbar\omega)^3\rho^2(\rho^2+1)$ up to terms associated with the zero-point energy, which do not modify the
large-\(\rho\) scaling. Using $\Delta\epsilon=-Mc^{2}\Delta\gamma$, which
follows from $\gamma_i=1-\epsilon_i/Mc^{2}$, the relative size of the nonlinear
correction reduces to $\frac{\delta V_{\Omega}}
{\delta V_{\rm TD}}
\sim
4\eta(\rho^2+1)$ where $\eta=
\frac{\hbar\omega}{Mc^2}$.
The correction becomes appreciable when $4\eta(\rho^2+1)\sim1$.
For large coherent amplitudes this condition reduces to $\rho^2\hbar\omega\sim Mc^2$.
Thus the nonlinear correction becomes significant only when the mean internal
energy of the oscillator clock approaches the rest-energy scale of the
composite system. The dominant observable effect
is expected to be the Gaussian visibility decay associated with ordinary
time dilation, while the \(\Omega(H_r)\)-induced correction should appear only
as a very small deviation from this Gaussian law. The resulting suppression and shift of the standard revivals is illustrated in Fig.~\ref{fig:visibility}, where it is shown the visibility $V$ of a coherent-state oscillator clock with mean excitation $\rho^{2}=\langle n\rangle=5$, as a function of the dimensionless time $\Delta\gamma\,\omega T$. In order to make
the effect of the normalization factor visible, the ratio $\eta=\hbar\omega
/Mc^{2}$ is set to the illustrative values $\eta=0.02$ and $\eta=0.05$, far
larger than any realistic atomic-clock value ($\eta\sim10^{-8}$); at physical
values the three curves would be indistinguishable on this scale. The dashed
curve ($\eta=0$) is the standard time-dilation result: because the oscillator
spectrum is equidistant, all energy components rephase simultaneously and the
visibility is periodic, decaying after the initial Gaussian falloff but
reviving completely at $\Delta\gamma\,\omega T=2\pi$ and $4\pi$. The solid
curves include the $\Omega(H_{r})$ correction, whose quadratic-in-energy term
adds an anharmonic phase $\propto\eta\,n^{2}$ to each component. This
anharmonicity spoils the exact rephasing, so the revivals are progressively
suppressed, shifted to later times, and broadened as $\eta$ increases, while
the short-time Gaussian decay---governed by ordinary time dilation---is
essentially unchanged. The deviation of the solid curves from the dashed one
is therefore a direct, if strongly magnified, signature of the nonlinear
internal-energy dependence introduced by $\Omega(H_{r})$.

The second example considered in this work was the hydrogen atom as the internal clock.
For hydrogenic transitions, the internal energy scale is of order eV, whereas
the rest energy is of order GeV. Hence $\frac{E_a}{Mc^2}\ll1$ and the correction induced by \(\Omega(H_r)\) is strongly suppressed. Nevertheless, hydrogen-like systems are conceptually useful because their small mass makes the ratio \(E_{\rm int}/Mc^2\) larger than in heavy atomic
clock systems. The normalization correction would appear as a small systematic
shift in the visibility oscillation frequency.

\subsection{Discussions}

The result above can be directly compared with the interferometric visibility
proposal of Zych et al.~\cite{zych}. In that work, a particle carrying an
internal clock travels in a superposition of two paths with different elapsed
proper times. The internal clock states become distinguishable, and the
visibility is given by the clock-state overlap $V=
|\langle \tau_1|\tau_2\rangle|$.
For a two-level internal clock initially prepared in an equal superposition of
energy eigenstates, this can be written as $V_{\rm Z}(T)=
\left|
\cos\left(
\frac{\Delta E\,\Delta\tau}{2\hbar}
\right)
\right|$
where \(\Delta E=E_1-E_0\) and \(\Delta\tau\) is the proper-time difference
between the two interferometric paths.

In this work we have shown that a normalization factor $\Omega$ appears in the center of mass system, then the corresponding two-level result is
\begin{equation}
V(T)=
\left|
\cos\left[
\frac{T}{2\hbar}
\left(
\Delta\gamma\,\Delta E
+
\Delta\epsilon\,\Delta\Omega
\right)
\right]
\right|,
\end{equation}
where $\Delta\Omega=\Omega(E_1)-\Omega(E_0)$.
Thus the usual proper-time result is recovered when the term
\(\Delta\epsilon\,\Delta\Omega\) is neglected. The new contribution induced by
\(\Omega(H_r)\) shifts the visibility phase by $\delta\theta_{\Omega}
=
\frac{T}{2\hbar}
\Delta\epsilon\,\Delta\Omega$.
For \(E/(Mc^2)\ll1\),
\begin{equation}
\Delta\Omega
\simeq
\frac{E_1^2-E_0^2}{M^2c^4}
=
\frac{\Delta E(E_1+E_0)}{M^2c^4}.
\end{equation}
Therefore $\delta\theta_{\Omega}
\simeq
\frac{T}{2\hbar}
\Delta\epsilon
\frac{\Delta E(E_1+E_0)}{M^2c^4}$. In a gravitational interferometer, the proper-time difference between two
arms at different heights is approximately $\Delta\tau
\simeq
\frac{\Delta V}{c^2}T$ where $\Delta V=g\Delta h$ \cite{roura2020}. The standard phase in the visibility is then $\theta_{\rm Z}
=
\frac{\Delta E\,\Delta\tau}{2\hbar}
\simeq
\frac{\Delta E}{2\hbar}
\frac{\Delta V}{c^2}T$.
If the external energy difference between the two branches is estimated as $\Delta\epsilon\simeq M\Delta V$, then the ratio between the new correction and the standard proper-time visibility phase is $\frac{\delta\theta_{\Omega}}{\theta_{\rm Z}}
\simeq
\frac{E_1+E_0}{Mc^2}$. Thus the correction is controlled by the ratio between the absolute internal energy scale of the clock and the rest energy of the composite system.
For heavy optical-clock atoms, this ratio is extremely small. For example,
for an optical transition with an energy scale of order eV and an atomic rest
energy of order \(10^{11}\,{\rm eV}\), one obtains $\frac{\delta\theta_{\Omega}}{\theta_{\rm Z}}
\sim
10^{-11}$.
For hydrogen, using \(Mc^2\simeq 0.94\,{\rm GeV}\) and an internal energy
scale of order \(10\,{\rm eV}\), one obtains instead $\frac{\delta\theta_{\Omega}}{\theta_{\rm Z}}\sim 10^{-8}$.
Hydrogen-like systems are therefore more favorable than heavy atoms, although
the correction remains very small.

The comparison with the quantum-clock analysis of Chiba and
Kinoshita~\cite{chiba} leads to a similar conclusion. In their estimates,
quantum time-dilation effects for atoms in momentum or position
superpositions can reach the scale $\Delta T_{\rm Q}\sim10^{-17}\,{\rm s}$ for experimental parameters. In the present model, the
normalization correction is suppressed by the same factor
\(E_{\rm int}/Mc^2\). For an optical transition in a heavy atom this gives $\Delta T_{\Omega}
\sim
10^{-11}\times10^{-17}\,{\rm s}
\sim
10^{-28}\,{\rm s}$. This is far below experimental achieved values \cite{chou2010,bothwell2022}. Hence, for current atomic-clock
platforms, the dominant observable effect is the standard time-dilation
visibility loss, while the \(\Omega(H_r)\) correction would appear only as a
tiny systematic deviation.
These estimates indicate that the most promising experimental strategy is not
necessarily to maximize only the transition energy \(\Delta E\), but rather to
compare transitions with the same \(\Delta E\) and different mean internal
energies. Indeed, the standard visibility phase depends on \(\Delta E\), while
the correction derived here depends on $E_1^2-E_0^2=\Delta E(E_1+E_0)$. Therefore, two internal clocks with the same transition frequency but
different values of \(E_1+E_0\) would have the same standard time-dilation
visibility but different \(\Omega(H_r)\)-induced corrections. A differential
visibility measurement of this type would help isolate the nonlinear
internal-energy dependence of the effective Hamiltonian.
Another possible route is to use very light systems, such as hydrogen,
positronium, or other systems with small rest energy, since the relevant
suppression factor is \(E_{\rm int}/Mc^2\). A more speculative possibility is
to use mesoscopic systems or optical cavities with a large coherent internal
energy \(E_{\rm int}=N\hbar\omega\) and a very small total mass. In such a
case the ratio $\frac{E_{\rm int}}{Mc^2}=\frac{N\hbar\omega}{Mc^2}$ could be enhanced by increasing the number of internal excitations. The main
experimental challenge would be to preserve the CM coherence while storing a
large coherent internal energy.
A concrete platform of this kind is the single-electron relativistic clock
interferometer proposed in Ref.~\cite{bushev}, where the clock is the spin
precession of an electron in a Penning trap and its time dilation depends on
the cyclotron (oscillator) state. The relativistically corrected cyclotron
ladder realizes precisely the anharmonic, energy-dependent spectrum
underlying the $\Omega(H_{r})$ correction discussed here, which makes it a
natural candidate for the differential visibility measurement proposed above.

The nonlinear correction discussed in this work may also be connected with
recent cold-atom analog experiments of relational time. In
Ref.~\cite{barontini2026}, a Bose-Einstein condensate is partitioned into an
observed bright sector and an unobserved dark sector. An internal entropic
time is constructed from the entropy flow of the bright sector, and the
measured evolution is reproduced by an effective Schr\"{o}dinger equation
parametrized by this internal time.

The effective equation derived in Ref.~\cite{barontini2026} follows from a
time-independent Wheeler-DeWitt-type constraint and contains a square-root
Hamiltonian of the form
\begin{equation}
i\hbar\partial_\tau\psi(\tau,a)
=
\frac{d\phi}{d\tau}
\sqrt{
\alpha^2\omega^2\phi^4
+
2\alpha\phi H_{\rm geom}
}
\,
\psi(\tau,a),
\end{equation}
where \(\phi\) plays the role of the clock variable and \(a\) the analog scale
factor. Expanding the square root gives
\begin{widetext}
\begin{equation}
i\hbar\partial_\tau\psi(\tau,a)
=
\Theta(\tau)\psi(\tau,a)
+
\Lambda(\tau)H_{\rm geom}\psi(\tau,a)
-
\Xi(\tau)H_{\rm geom}^2\psi(\tau,a)
+
\cdots .
\end{equation}
\end{widetext}
The first two terms reproduce the entropic-time Schr\"{o}dinger equation used in
Ref.~\cite{barontini2026}, while the term proportional to \(H_{\rm geom}^2\)
is the next-order correction. This structure is directly analogous to the correction obtained in the present work. Indeed, the normalization factor $\Omega(H_r)$ has the low-energy expansion
\begin{equation}
\Omega(H_r)
=
1+
\frac{H_r^2}{M^2c^4}
+
O\left(
\frac{H_r^3}{M^3c^6}
\right).
\end{equation}
Thus, in both cases the first nontrivial correction to the effective
relational dynamics is nonlinear in the Hamiltonian of the non-clock degrees
of freedom. This observation suggests that the cold-atom platform of
Ref.~\cite{barontini2026} could be useful to test the type of
nonlinear Hamiltonian corrections discussed in this work. In that setting, the
observable signature would not be an interferometric visibility loss, but
rather a systematic deviation of the measured bright-sector width
\(\Sigma(\tau)\) from the prediction of the first-order entropic Schr\"{o}dinger
equation. A simple estimate of the correction follows from the square-root expansion. If the clock contribution under the square root dominates the geometric contribution by a factor of order \(N_{\rm bright}\), then the relative size of the quadratic
correction is $\frac{\delta_{\rm quad}}{\delta_{\rm lin}}
\sim\frac{1}{4N_{\rm bright}}$. For \(N_{\rm bright}\sim10^4\), this gives a correction of order \(10^{-5}\),
which is below the current experimental uncertainty. However, the correction
is enhanced for smaller bright-sector populations and near the turning points
of the relational evolution, where the coefficient of the quadratic term grows
strongly. Another, more speculative, approach is to implement a Loschmidt-echo protocol by comparing two relational evolutions generated with different barrier heights \cite{Gorin2006}. The nonlinear term would produce an additional dephasing contribution to the echo amplitude, controlled by the difference between the corresponding quadratic coefficients. Such a protocol would provide a direct analog probe of the nonlinear Hamiltonian dependence emphasized in the present work.

\section{Conclusions}
\label{sec:conc}

In this work we have performed a detailed study of the emergence of
relativistic temporal effects in a composite quantum clock within the
Page--Wootters relational formalism. By considering a system with internal and
center-of-mass degrees of freedom and conditioning the global stationary state
on the internal clock, we have derived a Schr\"{o}dinger-type equation for the
center-of-mass wave function in the clock representation. By retaining the full
mass-energy structure of the composite particle, we have shown that the
conditioned center-of-mass Hamiltonian is multiplied by an operator-valued
factor $\Omega(H_{r})$ depending on the internal energy, which dresses the
center-of-mass energy with the internal energy and, being non-diagonal in the
clock-time basis, renders the conditioned equation non-local in the clock time
over a scale set by the Compton time $\hbar/Mc^{2}$. We have then applied this
framework to a center-of-mass momentum superposition, computing the
interferometric visibility with respect to the laboratory time, and we have
shown that $\Omega(H_{r})$ adds to the visibility a correction quadratic in the
internal energy---the linear term cancelling in the small-energy expansion---so
that the correction is sensitive to the absolute distribution of clock energies
and not only to the energy gaps that govern the standard time-dilation
dephasing. In turn, we have shown that the nonlocal clock-time equation and the
diagonal evolution in the internal-energy basis are two representations of the
same dynamics, and that the usual time-dilation visibility is recovered when
$\Omega(H_{r})\rightarrow I_{r}$.

\begin{acknowledgments}
This paper was partially supported by grant of Universidad Nacional del Sur (UNS)
through PGI 24/F094 F049PICT 1770. J. S. A. is a
member of CONICET.
\end{acknowledgments}

\appendix

\section{Coherent-state representation of the clock operator}

Coherent states are temporally stable under the harmonic oscillator evolution: their radius \(\rho\) is preserved, while their phase rotates with angular frequency \(\omega\). We now introduce a clock operator \(T\) satisfying formally $ [T,H_r]=i\hbar I_r$.  Equivalently, in the Heisenberg picture, $e^{iH_rt/\hbar}T e^{-iH_rt/\hbar} = T+tI_r$. We look for a coherent-state representation of \(T\) of the form
\begin{equation}
T= \frac{1}{\pi} \int_0^\infty \rho\,d\rho \int d\phi\, \mathcal T(\rho,\phi) |\rho,\phi\rangle\langle\rho,\phi| ,
\label{app:T-general-P}
\end{equation}
where we have used the resolution of the identity of Eq.~(\ref{0.4}).
The integral over \(\phi\) is understood over a chosen branch, for instance \(\phi\in[0,2\pi)\), with the usual caveat that a phase operator is globally multi-valued on the circle. Using $e^{iH_rt/\hbar} |\rho,\phi\rangle = |\rho,\phi+\omega t\rangle$, we have
\begin{equation}
e^{iH_rt/\hbar}T e^{-iH_rt/\hbar} = \frac{1}{\pi} \int_0^\infty \rho\,d\rho \int d\phi'\, \mathcal T(\rho,\phi'-\omega t) |\rho,\phi'\rangle\langle\rho,\phi'| 
\end{equation}
where we have used that $\phi'=\phi+\omega t $.
From the expressions above, the symbol \(\mathcal T(\rho,\phi)\) must satisfy $\mathcal T(\rho,\phi-\omega t) = \mathcal T(\rho,\phi)+t$. A solution is $\mathcal T(\rho,\phi) = -\frac{\phi}{\omega}$. Therefore the clock operator can be represented as given in Eq.~(\ref{0.6}). Finally, in terms of the complex variable $ z=\rho e^{i\phi}$ the phase can be written, on a chosen branch, as $\phi = -i\ln z+i\ln|z|$. Equivalently, $\phi=\arg z$. Thus the time operator can be written compactly as
\begin{equation}
T= -\frac{1}{\pi\omega} \int d^2z\, \arg(z)\, |z\rangle\langle z| ,
\label{app:T-complex}
\end{equation}
where $d^2z=\rho\,d\rho\,d\phi$. In this sense, the coherent-state symbol associated with the clock operator is $\mathcal T(z,z^*) = \frac{i}{\omega}\ln z - \frac{i}{\omega}\ln|z|$ again with the understanding that a branch of the logarithm must be fixed. In the Bargmann representation, one may also write $a^\dagger\rightarrow z$ and $a\rightarrow \frac{\partial}{\partial z}$ so that $\left[ \frac{\partial}{\partial z},z \right] = I$.   This representation makes explicit the canonical structure underlying the coherent-state construction, although the phase variable remains globally defined only modulo \(2\pi\). A more general description of the generalized coherent states representation can be found in \cite{foti}.

\end{document}